\documentclass[twocolumn,showpacs,preprintnumbers]{revtex4}

\usepackage{amsmath,amssymb,amsfonts}
\usepackage{bm,graphicx,color}
\usepackage{epsfig}
\usepackage{amsmath}
\newcommand{\laco}[1]{$\mathrm{LaCoO_3}$}

\newcommand{\lsrho}{(Sr$_{1-x}$La$_x$)$_2$RhO$_4$}
\newcommand{\sro}[1]{Sr$_2$RhO$_4$}

\begin{document}
\title{Doping-dependent bandwidth renormalization and spin-orbit coupling 
 in \lsrho }
\author{Kyo-Hoon Ahn$^1$}
\author{Kwan-Woo Lee$^{1,2}$}
\email{mckwan@korea.ac.kr}
\author{Jan Kune\v{s}$^3$}
\email{kunes@fzu.cz}
\affiliation{
$^1$Department of Applied Physics, Graduate School, Korea University, Sejong 339-700, Korea\\
$^2$Department of Display and Semiconductor Physics, Korea University, Sejong 339-700, Korea\\
$^3$Institute of Physics, Academy of Sciences of the Czech republic, Cukrovarnick\'a 10,
Praha 6, 162 53, Czech Republic}
\date{\today}

\begin{abstract}
We investigate the electronic structure of \lsrho  using a combination of the density functional 
and dynamical mean-field theories. Unlike the earlier 
local density approximation plus Hubbard $U$ (LDA+U) studies, 
we find no sizable enhancement of the spin-orbit splitting due to electronic correlations 
and show that such an enhancement is 
a spurious effect of the static mean-field approximation of the LDA+U method.
The electron doping suppresses the importance of electronic correlations, 
which is reflected in quasi-particle bandwidth increasing with $x$. \lsrho 
can be classified as weakly correlated metal, which becomes an itinerant in-plane ferromagnet 
(but possibly A-type antiferromagnet) due to Stoner instability around $x=0.2$.
\end{abstract}
\maketitle

\section{Introduction}
Metal-insulator transitions and the physics of strong electronic correlation 
in general have traditionally been studied in oxides of $3d$ elements and related compounds. 
The recent discoveries of spin-orbit coupling (SOC) assisted Mott state 
in Ba$_2$NaOsO$_6$\cite{bnoo07,bnoo14} and Sr$_2$IrO$_4$\cite{jyu08}, 
their potential similarity to cuprates, as well as the possibility of exotic phases
in iridates with honeycomb lattice\cite{jan09}, have attracted considerable attention 
to physics of oxides of transition metals with strong spin-orbit coupling. 

A decade ago, Perry {\it et al.}~\cite{perry06}  synthesized \sro~, which is isostructural 
to the unconventional superconductor Sr$_2$RuO$_4$.
Although there is only one more $4d$ electron in \sro~,
the measured electronic structures are significantly different 
from those of Sr$_2$RuO$_4$.\cite{perry06,baum06,ckim06} 
In Sr$_2$RuO$_4$, a good agreement is found 
between the first-principles calculations of the local density approximation (LDA)
and the experimental observations\cite{dama00}. 
In contrast, \sro~ was pointed out \cite{ckim06} as a curious example of a material
with relatively weak Coulomb interaction, where LDA fails 
to reproduce the experimental observations of the de Haas--van Alphen (dHvA) effect \cite{baum06} 
and angle resolved photoemission (ARPES).\cite{ckim06} 
Moreover, it was shown~\cite{liu08,haver08} that SOC affects the shape of the Fermi surface
substantially.
Using the LDA+U calculations,
Liu {\it et al.}~\cite{liu08} argued that on-site Coulomb interaction $U$ effectively enhances the 
strength of SOC. 
Later LDA plus the dynamical mean field theory (DMFT) calculations by Martins {\it et al.}~\cite{martins} 
did not find such an enhancement, but they did not analyze this aspect of their results. 

Furthermore, it has been debated whether a replacement of some of Sr by La ions, 
i.e., electron doping to the $4d$ Rh bands, 
leads to a long-range magnetic order in this system.\cite{furuta}
Recently, Kim {\it et al.} synthesized single crystalline \lsrho~ samples
up to $x=0.25$.\cite{chul1,chul2} 
Above $x=0.2$, the susceptibility shows a kink at $T_N=20$ K,
implying a magnetic ordering.
However, neither detailed experimental nor theoretical investigation 
has yet been performed.

Here, we report the results of combined LDA and DMFT calculations, including SOC, 
and investigate the effect of electronic correlations in stoichiometric and electron doped \sro. 
The computations follow the approach used in Ref.~[\onlinecite{arita12}],
to study an isoelectronic compound Sr$_2$IrO$_4$.

\section{Numerical method}
The LDA+DMFT calculations proceed in two steps.\cite{dmft,held06} 
First, we carried out density functional calculations in LDA,
including SOC, using the {\sc wien2k} package.\cite{wien2k,wien2k2}
We used the experimental crystal structure of
the K$_2$NiF$_4$ type (space group: $I4_1/acd$, No. 142) with the lattice constants
$a=5.43562$ \AA~ and $c=25.77490$ \AA.\cite{str1}
The La doping was treated in a virtual crystal approximation. 
In the doped cases, the internal parameters of Sr and O ions were optimized, 
while the lattice constants were kept fixed to their stoichiometric values.
This is justified by
no significant change of the lattice constants being observed
in experiment.\cite{chul1}
The basis size was determined by $R_{mt}K_{max}=7$ with atomic radii in bohr:
Sr (2.31), Rh (1.97), and O (1.70).

Next, Wannier functions spanning the $t_{2g}$ manifold and the corresponding hopping amplitudes 
were obtained using the {\sc wannier90}~\cite{marzari} and {\sc wien2wannier}~\cite{jan10} 
programs. The choice of the  $t_{2g}$ manifold as the one-particle basis of our model is justified
{\it a posteriori} by an observation of minor spectral-weight transfer to Hubbard bands and the fact
that the $t_{2g}$ quasi-particle band is well separated from the O-$p$ band.
The $p$-$t_{2g}$ mixing in the correlated model is therefore similar to that of LDA, implicitly fixed
by the Wannier construction.
Construction of the effective Hubbard Hamiltonian was concluded by adding
the intra-atomic repulsion $H_{\text{int}}$, parametrized by $F^0=1.6$~eV and $J=0.3$~eV taken from Ref.~[\onlinecite{martins}].
Only the density-density terms in the $|j,j_z\rangle$ basis, $H_{\text{int}}=\sum_{i,\alpha,\beta}U_{ij}n^{\alpha}_in^{\beta}_i$, were kept in the actual calculations (for the numerical values of $U_{ij}$ see Appendix).
We have employed the segment implementation of the hybridisation-expansion 
continuous-time quantum Monte-Carlo (QMC) algorithm~\cite{werner} 
to solve the auxiliary Anderson impurity problem. To obtain 
the one-particle spectra the self-energies were measured
using the equation of motion technique~\cite{hafer} combined 
with the improved estimators~\cite{augustinsky} and analytically continued
with the maximum-entropy method~\cite{maxent}.

The LDA+U calculations,\cite{ldau1,ldau2} performed for purpose of comparison, 
employed the standard {\sc wien2k} implemention 
with the fully-localized-limit double-counting correction.\cite{fll} 
In these calculations the same interaction parameters, $U=2.5$ eV and $J=0.9$ eV, as in Ref.[\onlinecite{liu08}],
were used. 
We point out that different interaction parameters used in LDA+U and LDA+DMFT calculations
are not necessarily inconsistent. Smaller interaction strength in DMFT reflects the use
of a smaller, $t_{2g}$-only, effective Hamiltonian, for which the interaction
is screened more strongly than for Hamiltonians including all Rh $4d$ bands and O $2p$ states.~\cite{screen1}

\section{Results and discussion}

\begin{figure}[tbp]
{\resizebox{8cm}{6cm}{\includegraphics{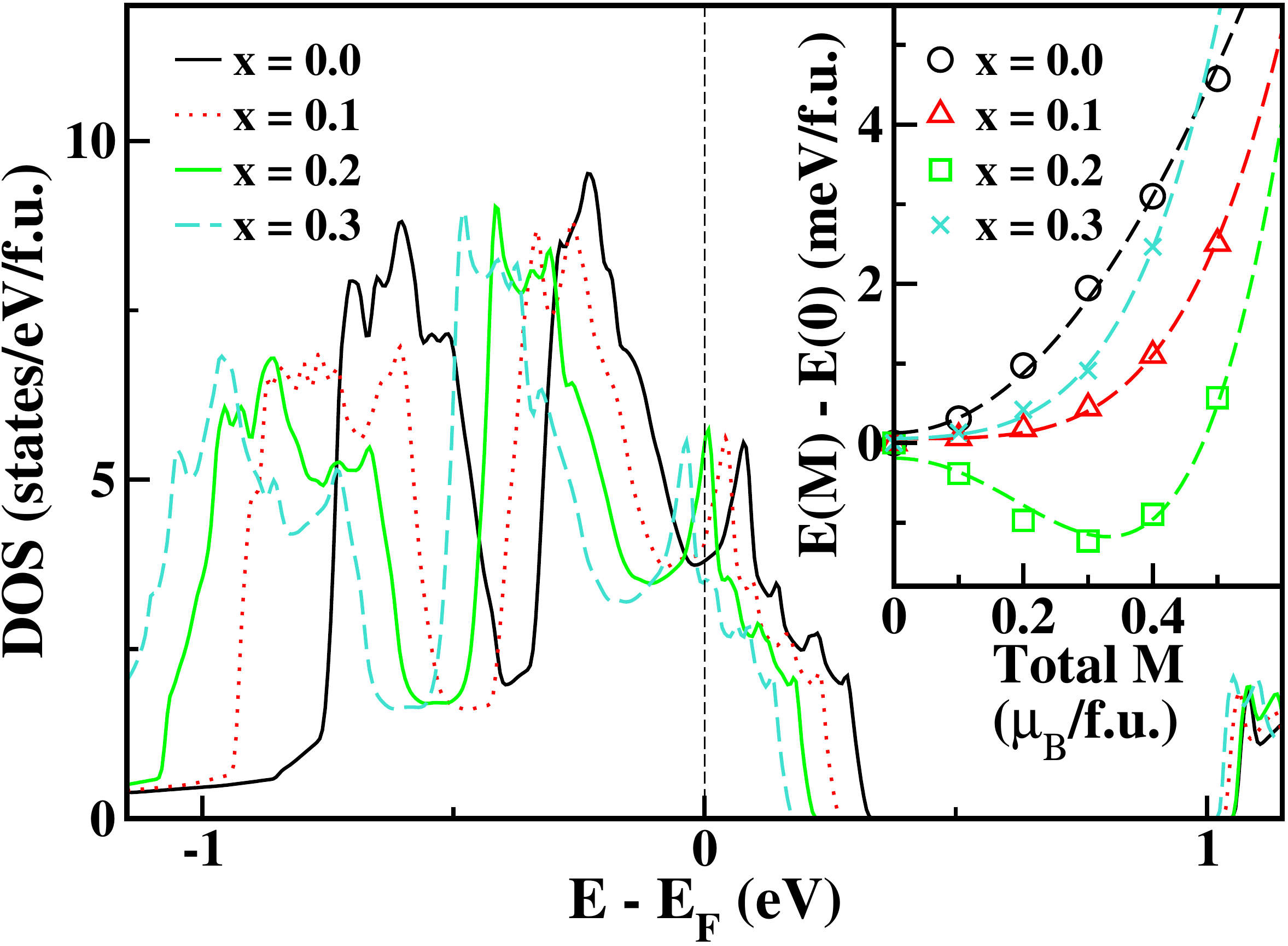}}}
\caption{(Color online) Variation of the LDA density of states (spectral density) for $x=$0--0.3.
For $x=0.2$, a sharp peak appears the Fermi level $E_F$.
Inset: The variation of the LDA total energy $\Delta E(M)=E(M)-E(0)$ versus
the ferromagnetic polarization. The dashed lines are guides for the reader's eye.
}
\label{fig:1}
\end{figure}

\subsection{Electron-doping dependent magnetic instability}
Before addressing the effect of dynamical correlation, 
we briefly investigate the possibility of magnetic ordering for La doping, 
i.e., electron doping,
using LDA and fixed spin moment calculations (FSM).\cite{krasko1987}

We start the presentation of our results with a plain density functional electronic structure. 
Figure \ref{fig:1} shows the LDA spectral densities obtained for La doping of $x=$0--0.3. 
The Fermi level $E_F$ lies inside the Rh $t_{2g}$ manifold, 
which is separated by a crystal-field gap from the Rh $e_g$ states above.
The electron doping causes essentially a rigid band shift, which moves
the peak arising from the van Hove singularity at the $X$ point
towards the Fermi level. The spectra obtained with LDA+U and LDA+DMFT, which we discuss later,
exhibit similar behavior.
At $x\approx0.2$ the Fermi energy coincides with the maximum
of the spectral density giving possibly rise to a Stoner instability.

To explore this possibility, we have performed FSM calculations in the range of $x=$0--0.3.
The resulting energy vs magnetization curves $E(M)$ are shown in the inset of Fig. \ref{fig:1}.
At $x=$0.2 a shallow energy minimum ($\sim$ 1.2 meV/f.u. below the nonmagnetic state) 
occurs for ordered moment of $M\approx$0.35 $\mu_B$/f.u.,
while nonmagnetic ground states are found for the other doping.
The LDA therefore predicts magnetic instability 
when the van Hove singularity is very close to $E_F$.
The Stoner effective exchange parameter $I$ can be obtained from the curvature of $E(M)$ 
at small $M$ regime
using the relations $E(M)=\tfrac{1}{2}\chi^{-1}M^2$ and $\chi=2\mu_B^2N(E_F)/(1-N(E_F)I)$,
with $N(E_F)$ being the density of states per spin channel at $E_F$. 
The obtained Stoner interaction is $I$=0.45 eV for $x$=0.2, and
$IN(E_F)\approx1.27$, above the Stoner criterion. 
This indicates that the system is weakly magnetically unstable 
in a narrow regime around $x=0.2$, 
in a good agreement with the recent experimental observation.\cite{chul1,chul2}

A consistent picture is provided by LDA+U calculations for $x=0.2$, where 
we obtained stable ferromagnetic (FM) and A-type antiferromagnetic (AFM) solutions 
with $M\approx 0.35 \mu_B$/Rh differing by less than 0.1 meV, 
but no stable solution with in-plane AFM order.

\begin{figure}[tbp]
\includegraphics[scale=0.3]{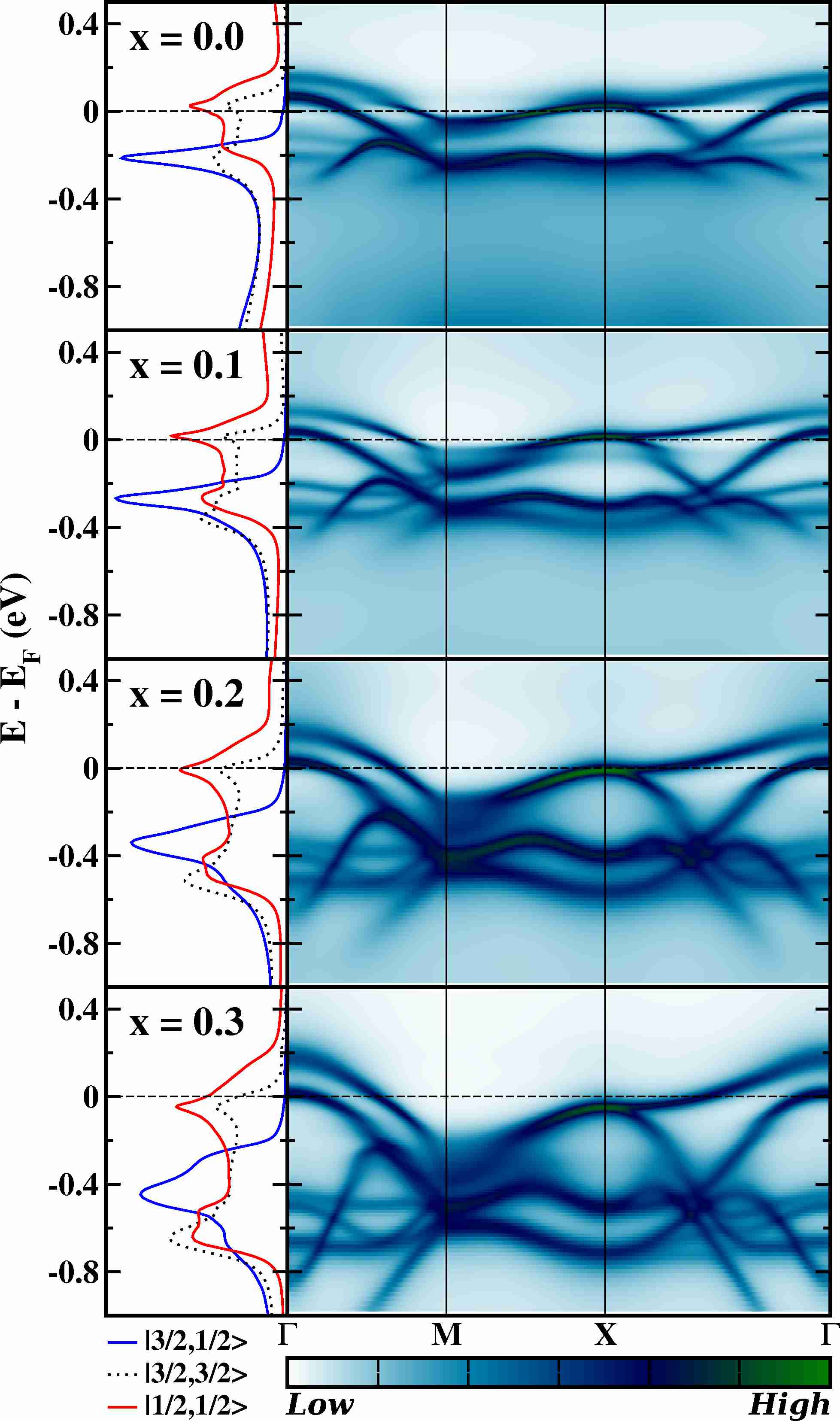}
\caption{(Color online) Spectral densities (left) and band structures (right)
of \lsrho~ at $x=$0, 0.1, 0.2, 0.3 (from top to bottom). 
}
\label{fig:2}
\end{figure}

\begin{figure}[tbp]
\includegraphics[scale=0.3]{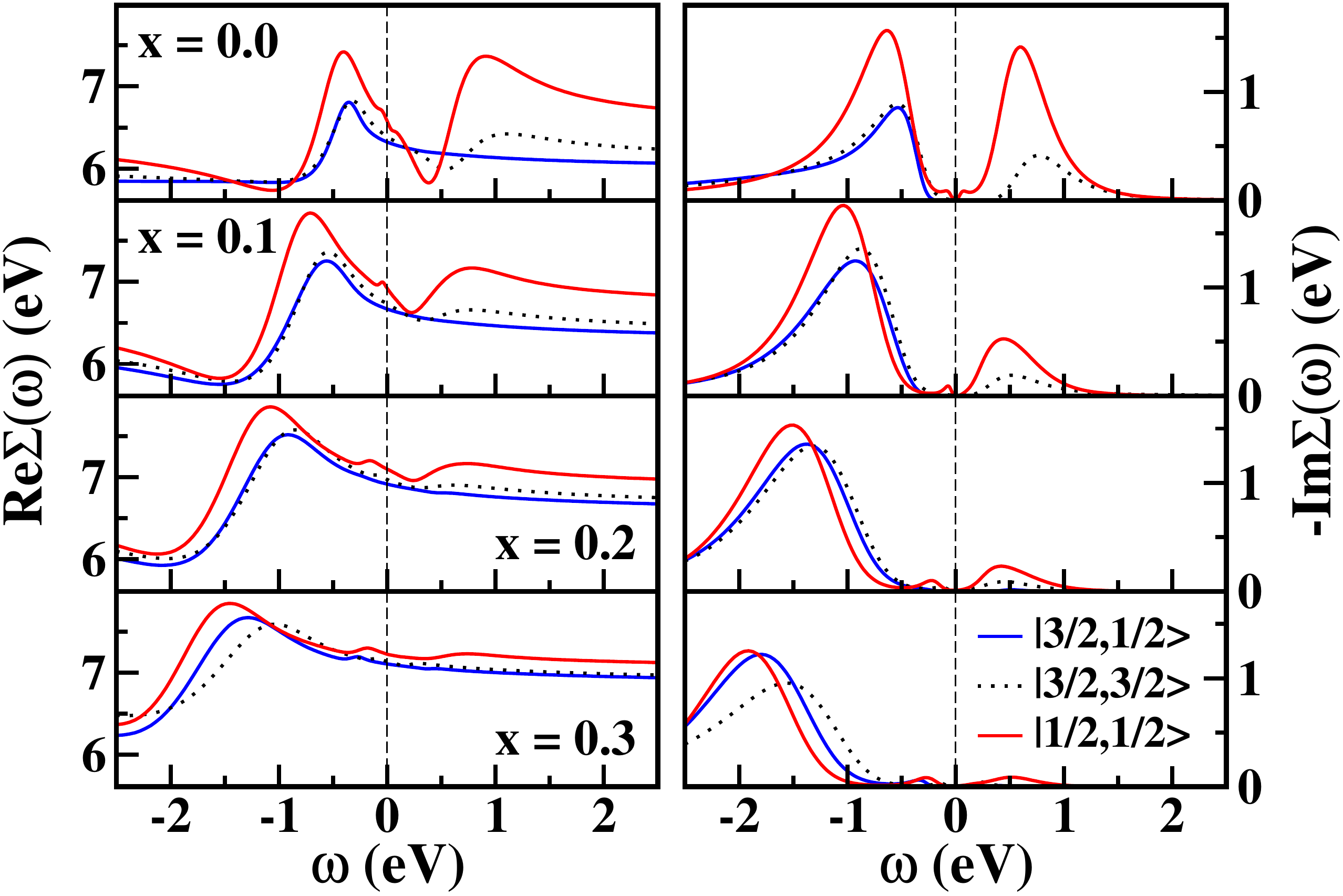}
\caption{(Color online) Frequency dependent real (left) and imaginary (right) 
parts of the self-energies $\Sigma(\omega)$ 
of \lsrho~ at $x=$0, 0.1, 0.2, 0.3 (from top to bottom). 
}
\label{fig:3}
\end{figure}

\subsection{Effects of dynamical correlation: LDA+DMFT approach}
The main effect of the local Coulomb interaction at low carrier density ($\leq1$ in the present case)
consists in reduction of the charge fluctuations.
This, in general, leads to a transfer of spectral weight to incoherent
features on the energy scale set by $U$ 
and renormalization of the quasi-particle (QP) bands. 
Figure \ref{fig:2} shows the orbital, $A(\omega)$, and k-resolved,  $A_{\mathbf{k}}(\omega)$, 
spectral densities for La doping levels between $x=$0--0.3. 
The undoped data essentially reproduce the results of Martins {\it et al.}~\cite{martins}.
Narrowing of the QP bands to their LDA counterparts is apparent. 
One can also distinguish the overall band narrowing
from even larger effective mass enhancement in the narrow Fermi liquid regime around $E_F$. 
While the kinks in the QP dispersion~\cite{byczuk} marking the Fermi liquid regime 
can hardly be resolved in the $A_{\mathbf{k}}(\omega)$, 
they are clearly visible in the self-energies $\Sigma(\omega)$ shown in Fig.~\ref{fig:3}.

\begin{figure}[tbp]
{\resizebox{8cm}{8cm}{\includegraphics{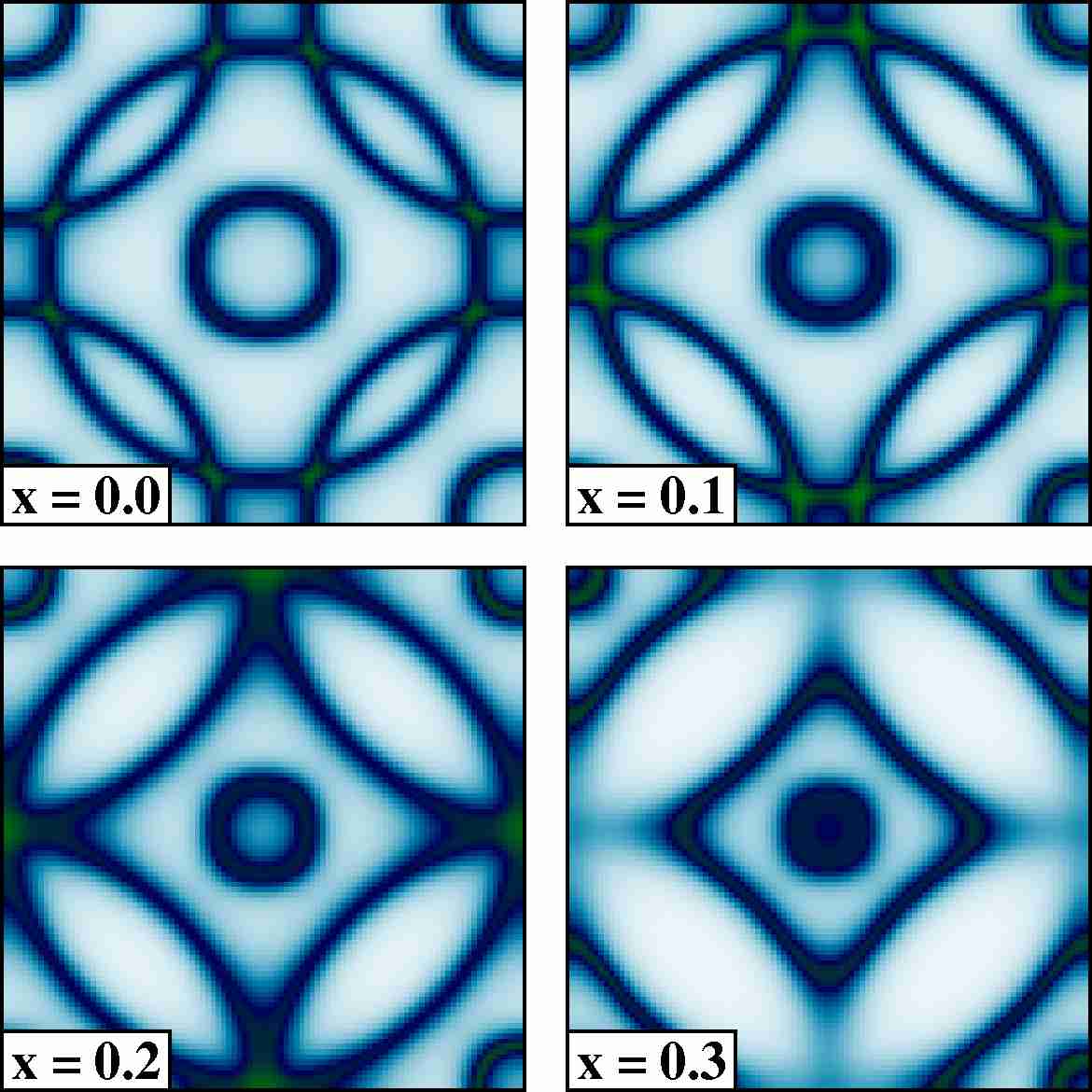}}}
\caption{(Color online) 2-dimensional Fermi surfaces on the $k_z=0$ plane 
of \lsrho~ at $x=$0--0.3.
The $\bar{\Gamma}$ and $\bar{M}$ points are at the center and each corner, respectively.
}
\label{fig:4}
\end{figure}

\begin{figure}[tbp]
\includegraphics[width=0.9\columnwidth]{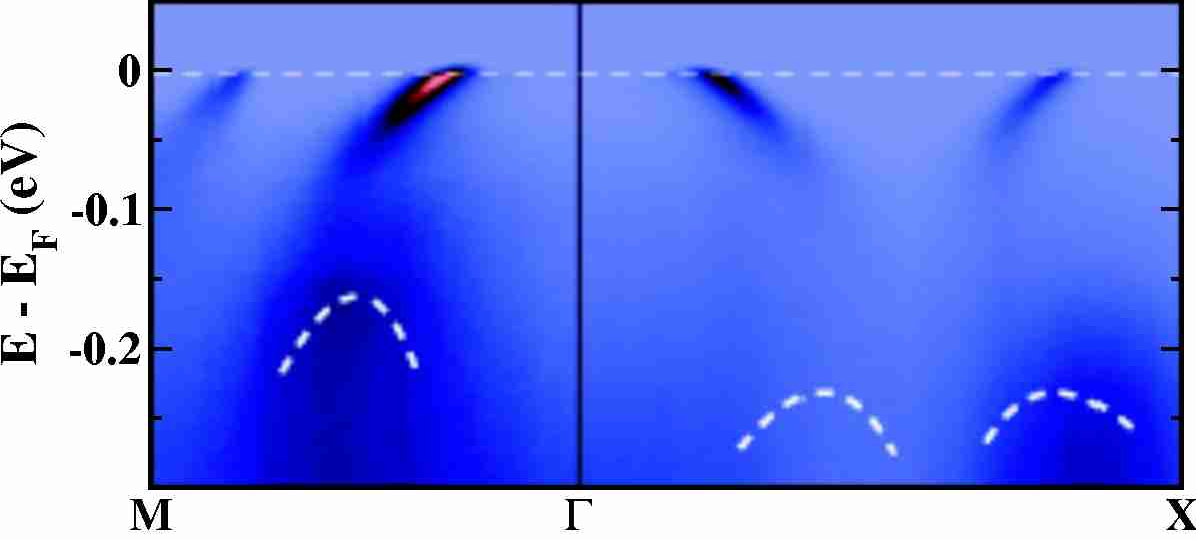}
\includegraphics[width=0.9\columnwidth]{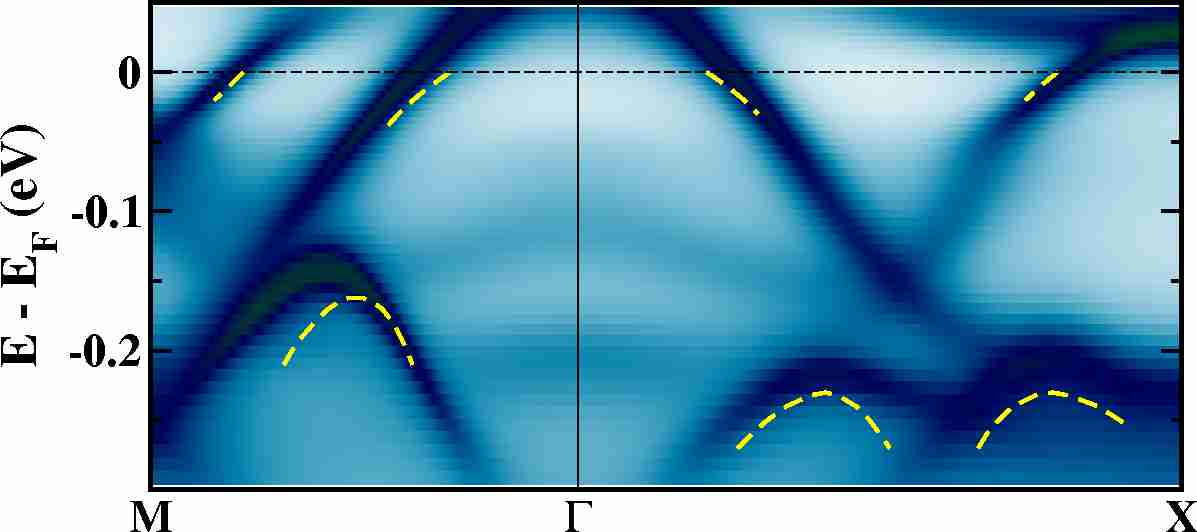}
\caption{(Color online) Detail of the QP bands in the vicinity of the Fermi level.
Top: the ARPES data taken with permission of authors from Ref.~\cite{perry06}. 
Bottom: the DMFT QP bands the $M-\Gamma-X$ line near $E_F$ at $x=0$. The dashed lines
mark the position and the slope of the ARPES bands from the top panel. 
}
\label{fig:5}
\end{figure}

\begin{figure}[tbp]
\includegraphics[width=0.9\columnwidth]{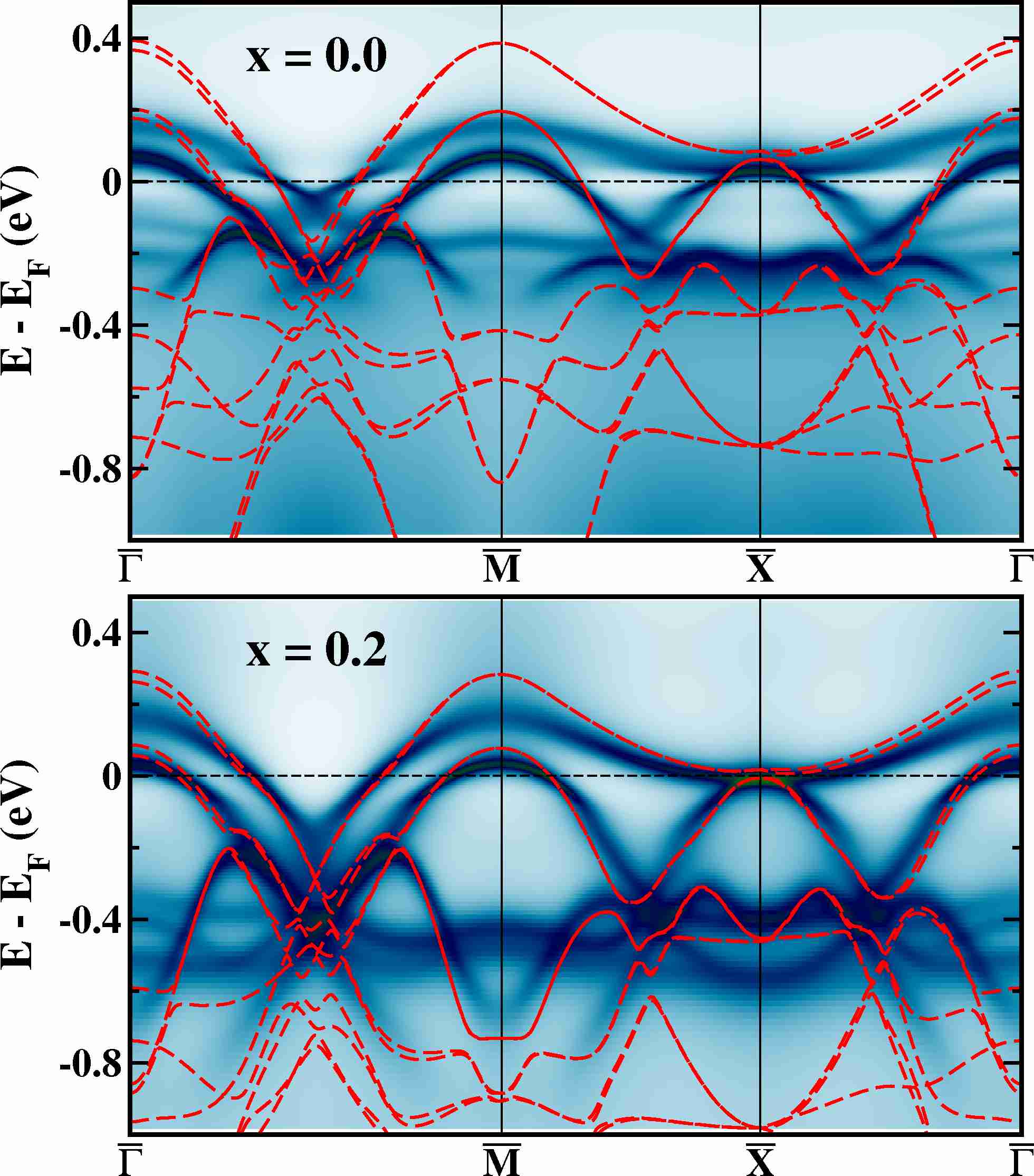}
\caption{(Color online) Comparison of LSDA+SO+U (red dashed lines) and
DMFT band structures at $x=$0 (top) and 0.2 (bottom). 
}
\label{fig:6}
\end{figure}

The theoretical interest in Sr$_2$RhO$_4$ was initiated by the unusually large discrepancy 
between the experimental Fermi surface measured by ARPES~\cite{perry06} and the results of
LDA calculations, as mentioned in the Introduction. 
Liu~{\it et al.}~\cite{liu08} showed that by including 
the local interaction on the Hartree-Fock level (LDA+U) an almost perfect match was obtained. 
They argued that the improved description of the Fermi surface is to
be attributed to the Coulomb enhancement of SOC. This enhancement is manifested
as increased spin-orbit splitting, e.g., at the $\bar{\Gamma}$ or $\bar{M}$ points of the Brillouin zone.
Here we show that this enhancement of the spin-orbit splitting is a spurious effect of the static
mean-field approximation of LDA+U.
The DMFT Fermi surfaces for various doping levels are displayed in Fig.~\ref{fig:4}.
Like LDA+U,\cite{liu08} 
the DMFT QP bands reproduce the experimental Fermi surface very well, 
but do not exhibit any sizable enhancement of the spin-orbit splitting. 
Moreover, the DMFT bands capture not only the Fermi surface correctly, 
but also the bands dispersion shown in Fig.~\ref{fig:5}.
This is obviously not the case of LDA+U bands which exhibit 2--3 times larger Fermi velocities
(see Fig.~\ref{fig:6}).
These results are easy to understand. 
In a full many-body treatment, the intra-atomic Coulomb interaction does not enhance spin-orbit splitting,  
but effectively (on a low-energy scale) suppresses its competitors.
The (on-site) SOC approximately commutes with the intra-atomic Coulomb interaction
and thus the spin-orbit splitting is not affected by the intra-atomic Coulomb interaction.
However, the Coulomb interaction renormalizes the QP mass, which leads to reduced bandwidths. 
The effect of SOC, which depends on the relative size of spin-orbit splitting and the bandwidths, 
is therefore enhanced.
This is the same mechanism as enhancement of orbital polarization (effect of a crystal field) 
due to the intra-atomic Coulomb interaction~\cite{pavarini}. 
The mean-field approximation of LDA+U, on the other hand, does not affect the bandwidth, 
but enhances the spin-orbit splitting. 
It is clear from the band geometry in \sro~, given in Fig.~\ref{fig:6}, 
that the size of the Fermi surface pockets depends 
on the ratio of the spin-orbit splitting to the bandwidth. Since the LDA+U gets this
ratio correctly, albeit for a wrong reason, it yields the correct Fermi surface. 
The deficiency of the LDA+U description becomes apparent when not only Fermi surface, 
but the full band dispersion is considered.

\begin{table}[bt]
\caption{Orbital resolved mass enhancements $Z^{-1}$
for various doping levels.
}
\begin{center}
\begin{tabular}{ccccc}\hline\hline
  La-doping~x~ &~~ $|3/2,1/2\rangle$~~ & ~~$|3/2,3/2\rangle$~~ &~~ $|1/2,1/2\rangle$~~ \\\hline
    0 &  1.53 & 1.92 & 4.02  \\
    0.1 & 1.46 & 1.83 & 2.90  \\
    0.2 & 1.32 & 1.66 & 1.61  \\
    0.3 & 1.12 & 1.36 & 1.42  \\\hline\hline
\end{tabular}
\end{center}
\label{tab:1}
\end{table}

\begin{figure}[tbp]
\vskip 8mm
\includegraphics[scale=0.3]{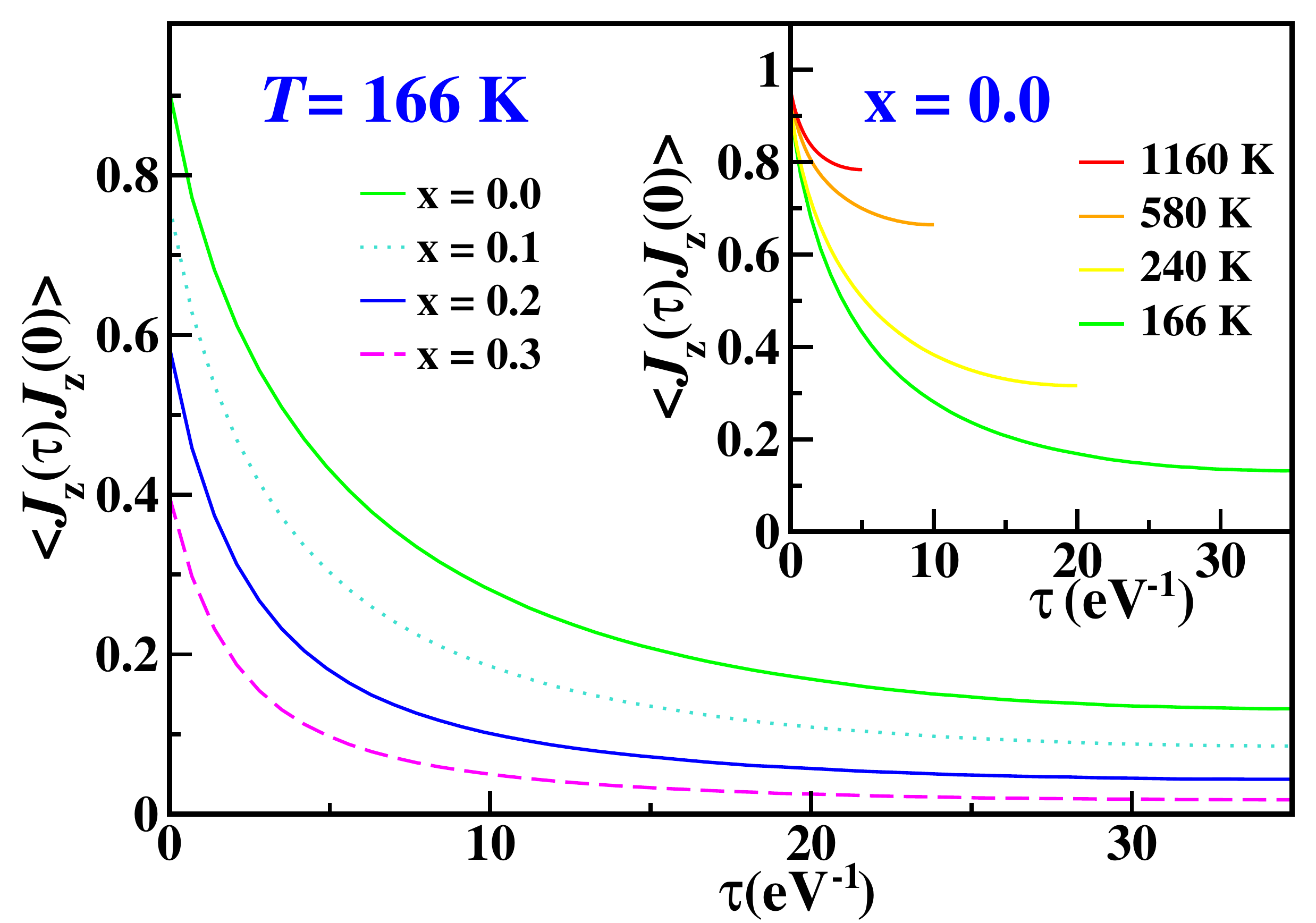}
\caption{(Color online) Imaginary-time local spin-spin correlation function
$\langle J_z(\tau)J_z(0)\rangle$ for various La-doping levels at $T=166$~K.
Inset: the spin-spin correlation functions at various temperatures $T$ for the undoped system.\\
}
\label{fig:7}
\end{figure}

With electron doping the charge carriers (holes) become rather dilute making the correlation effects
progressively weaker. The mass enhancement diminishes and the QP bands approach their 
uncorrelated (LDA) widths. This effect is quantified in Table~\ref{tab:1}, 
which shows the mass enhancement 
$Z^{-1}=\frac{m^\ast}{m}=(1-\frac{\partial Re\Sigma(\omega)}{\partial \omega})\mid_{\omega=0}$ 
for each orbital as a function of doping. 
Most pronounced variation is found for the $|\frac{1}{2},\frac{1}{2}\rangle$ 
orbitals which have the largest contribution to QP bands at $E_F$. Another important 
quantity connected to the correlation strength is the spin-spin correlation function, 
which can be used to detect fluctuating local moments. 
Figure \ref{fig:7} shows the imaginary-time local moment correlation functions $\langle J_z(\tau)J_z(0)\rangle$
for various doping levels. The rapid decay with $\tau$ shows that there are no long lived
local moments and that the magnetic ordering at $x=0.2$ arises 
as a weak-coupling Fermi surface instability.
This is consistent with the behavior of magnetically ordered solutions in LDA and LDA+U. In particular,
the fact that a stable FM solution exists, but in-plane AFM solution can not be stabilized.

\section{Conclusions}
Using the LDA+DMFT approach we have studied the effects of SOC and electronic correlations
in stoichiometric and electron doped \lsrho~. As observed previously, 
it is important to include both SOC and the on-site Coulomb interaction in order to reproduce 
the experimental Fermi surface of \sro~. However, in contrast to the claims of correlation
enhanced spin-orbit splitting throughout LDA+U approach~\cite{liu08}, 
our results do not show such enhancement. 
Instead we find that spin-orbit splitting is not affected by electronic correlations. 
It is its ratio to the bandwidth
that increases due to bandwidth renormalization, a behavior analogous to that of crystal-field splitting
observed in $d^1$ perovskites~\cite{pavarini}.

Our results suggest diminishing of electronic correlations with electron doping, 
leading to a sizable increase of the Fermi velocity that should be observable in ARPES experiments. 
Around the La doping of 0.2
the material becomes an itinerant in-plane ferromagnet with weak inter-plane coupling. 
There is no evidence  of fluctuating local moments in the paramagnetic state 
above the transition temperature.

\begin{acknowledgments}
We acknowledge Ch. Kim and C. Kim for providing their unpublished experimental data, 
M. Haverkort for useful discussion on a recent ARPES measurement, 
and R. Arita for a fruitful discussion.
K.W.L acknowledges the hospitality of the department of physics of Univ. of California, Davis,
during the preparation of this manuscript.
This research was supported by National Research Foundation of Korea Grant No. NRF-2013R1A1A2A10008946
(K.H.A and K.W.L.) and by Grant No. 13-25251S of the Grant Agency of the Czech Republic (J.K.).
\end{acknowledgments}

\begin{table}[bt]
\caption{Interaction matrix for the six orbits of the Rh $t_{2g}$ manifold (in units of meV), 
where 1, 2, 3, 4, 5, 6 denotes 
$|3/2,1/2\rangle$, $|3/2,-1/2\rangle$, $|3/2,3/2\rangle$, $|3/2,-3/2\rangle$, $|1/2,1/2\rangle$, 
and $|1/2,-1/2\rangle$, respectively.
Here, $F_0 = 1.6$ eV and $J = 0.3$ eV.
}
\begin{center}
\begin{tabular}{c|cccccc}\hline\hline
      \ & 1      & 2      & 3      & 4      & 5      & 6      \\\hline
    1 \ & 0.0000 & 1.7114 & 1.4027 & 1.4027 & 1.4799 & 1.4027 \\
    2 \ & 1.7114 & 0.0000 & 1.4027 & 1.4027 & 1.4027 & 1.4799 \\
    3 \ & 1.4027 & 1.4027 & 0.0000 & 1.7114 & 1.5570 & 1.3255 \\
    4 \ & 1.4027 & 1.4027 & 1.7114 & 0.0000 & 1.3255 & 1.5570 \\
    5 \ & 1.4799 & 1.4027 & 1.5570 & 1.3255 & 0.0000 & 1.6342 \\
    6 \ & 1.4027 & 1.4799 & 1.3255 & 1.5570 & 1.6342 & 0.0000 \\\hline\hline
\end{tabular}
\end{center}
\label{tab:2}
\end{table}

\begin{figure}[tbp]
\vskip 12mm
\includegraphics[width=\columnwidth,clip]{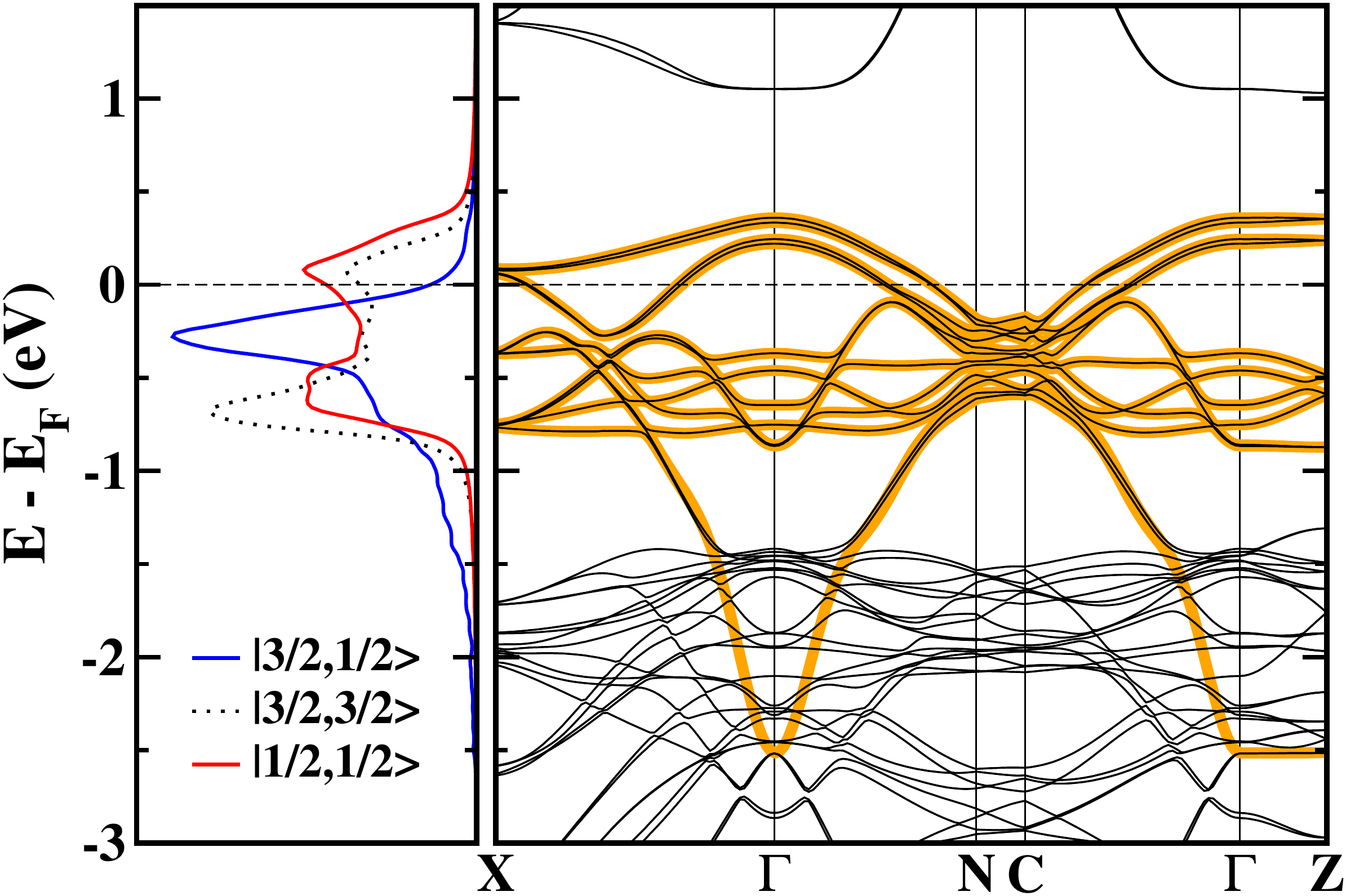}
\caption{(Color online) Our Wannier function fits (orange color) of the 24 LDA+SOC 
 bands of the $t_{2g}$ manifold in the cells containing 4 formula units. 
}
\label{fig:8}
\end{figure}

\begin{figure}[tbp]
\vskip 8 mm
\includegraphics[width=0.7\columnwidth,clip]{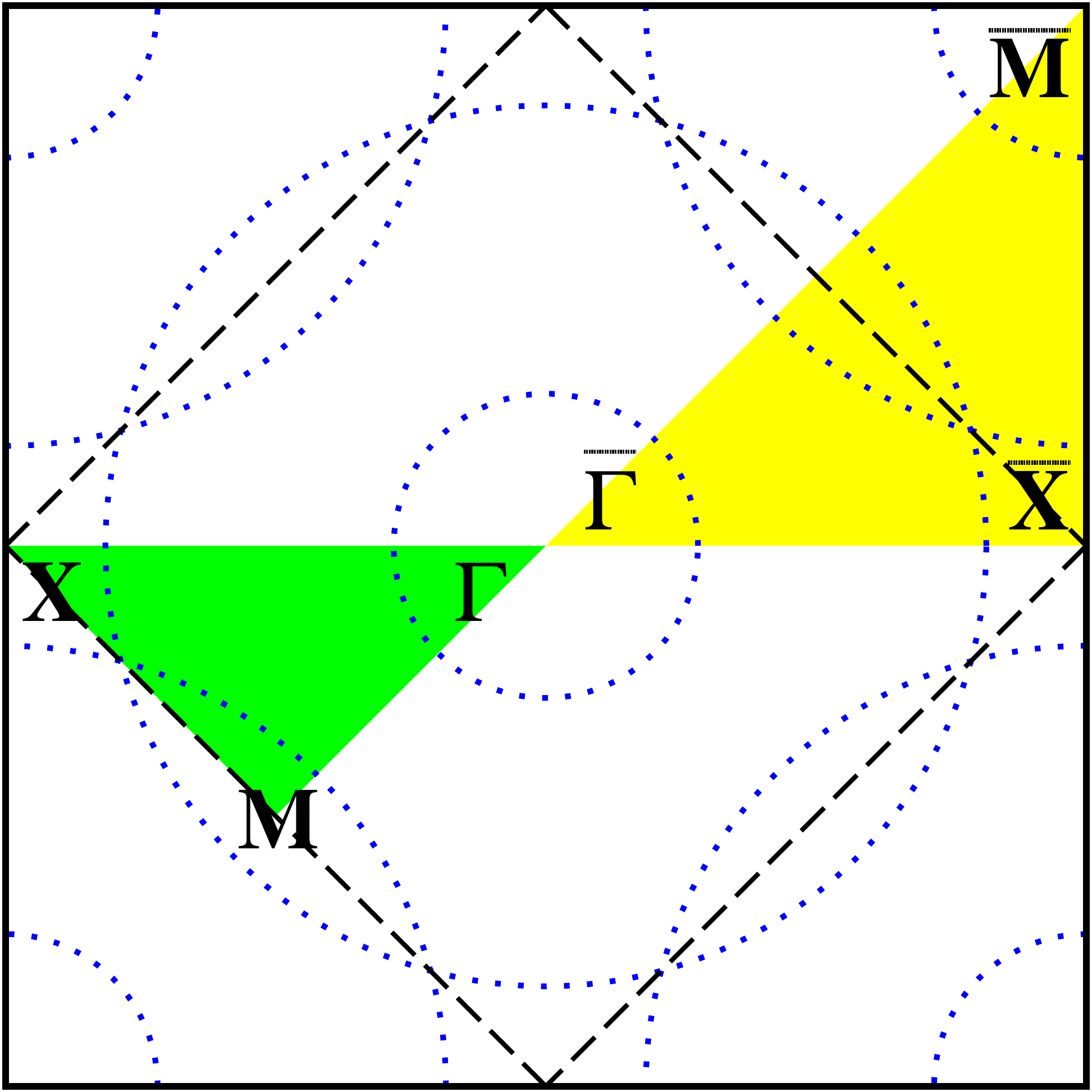}
\caption{(Color online) Comparison of high symmetry points
between 2-dimensional Brillouin zones of the orginal and doubled cells in \sro~.
}
\label{fig:9}
\end{figure}

\appendix
\section{}
Table \ref{tab:2} shows the interaction parameters (density-density) between the Rh $t_{2g}$ orbitals
in the basis $|j,m_j\rangle$.
In Fig. \ref{fig:8}, the LDA bands are compared to the bands arising from
Wannier functions spanning the $t_{2g}$ manifold. The high symmetry points used for the band structure plots
are shown in Fig.~\ref{fig:9}.


\begin{thebibliography}{34}
\bibitem{bnoo07} K.-W. Lee and W. E. Pickett,
 Europhys. Lett. {\bf 80}, 37008 (2007). 

\bibitem{bnoo14} S. Gangopadhyay and W. E. Pickett,
 arXiv:1410.3167.

\bibitem{jyu08} B. J. Kim, H. Jin, S. J. Moon, J. Y. Kim, B. G. Park, 
 C. S. Leem, J. Yu, T. W. Noh, C. Kim, S. J. Oh, J. H. Park,
 V. Durairaj, G. Cao, and E. Rotenberg, 
 Phys. Rev. Lett. {\bf 101}, 076402 (2008).

\bibitem{jan09} A. Shitade, H. Katsura, J. Kune\v{s}, X.-L. Qi, S.-C. Zhang, and N. Nagaosa,
 Phys. Rev. Lett. {\bf 102}, 256403 (2009).

\bibitem{perry06} R. S. Perry, F. Baumberger, L. Balicas, N. Kikugawa, N. J. C. Ingle, 
 A. Rost, J. F. Mercure, Y. Maeno, Z.-X. Shen, and A. P. Mackenzie, 
 New J. Phys. {\bf 8}, 175 (2006)

\bibitem{baum06} F. Baumberger, N. J. C. Ingle, W. Meevasana, K. M. Shen, D. H. Lu, 
 R. S. Perry, A. P. Mackenzie, Z. Hussain, D. J. Singh, and Z.-X. Shen, 
 Phys. Rev. Lett. {\bf 96}, 246402 (2006).

\bibitem{ckim06} B. J. Kim, J. Yu, H. Koh, I. Nagai, S. I. Ikeda, S.-J. Oh, and C. Kim, 
 Phys. Rev. Lett. {\bf 97}, 106401 (2006).


\bibitem{dama00} A. Damascelli, D. H. Lu, K. M. Shen, N. P. Armitage, F. Ronning, D. L. Feng, 
 C. Kim, Z.-X. Shen, T. Kimura, Y. Tokura, Z. Q. Mao, and Y. Maeno,
 Phys. Rev. Lett. {\bf 85}, 5194 (2000).


\bibitem{liu08} G.-Q. Liu, V. N. Antonov, O. Jepsen, and O. K. Andersen,
  Phys. Rev. Lett. {\bf 101}, 026408 (2008).

\bibitem{haver08} M. W. Haverkort, I. S. Elfimov, L. H. Tjeng, G. A. Sawatzky, 
 and A. Damascelli, 
 Phys. Rev. Lett. {\bf 101}, 026406 (2008).

\bibitem{martins} C. Martins, M. Aichhorn, L. Vaugier, and S. Biermann,
   Phys. Rev. Lett. {\bf 107}, 266404 (2011).


\bibitem{furuta} N. Furuta, S. Asai, T. Igarashi, R. Okazaki, Y. Yasui, I. Terasaki, 
 M. Ikeda, T. Fujita, M. Hagiwara, K. Kobayashi, R. Kumai, H. Nakao, and Y. Murakami,
 Phys. Rev. B {\bf 90}, 144402 (2014).


\bibitem{chul1} Ch. Kim, Ph. D. thesis, department of physics, Yeonsei Univ., Korea (2011).

\bibitem{chul2} C. Kim (private communications).


\bibitem{arita12} R. Arita, J. Kune\v{s}, A. V. Kozhevnikov, A. G. Eguiluz, and M. Imada, 
 Phys. Rev. Lett. {\bf 108}, 086403 (2012).


\bibitem{dmft} For review, see A. Georges, G. Kotliar, W. Krauth,
 and M. J. Rozenberg, Rev. Mod. Phys. {\bf 68}, 13 (1996); 
 G. Kotliar, S. Y. Savrasov, K. Haule, 
 V. S. Oudovenko, P. Parcollect, and C. A. Marianetti,
 {\it ibid.} {\bf 78}, 865 (2006).

\bibitem{held06} K. Held, I.A. Nekrasov, G. Keller, V. Eyert, N. Blumer, A.K. McMahan, 
 R.T. Scalettar, T. Pruschke, V.I. Anisimov, and D. Vollhardt,
 Phys. Status Solidi B {\bf 243}, 2599 (2006).


\bibitem{wien2k} K. Schwarz and P. Blaha,
 Comput. Mater. Sci. {\bf 28}, 259 (2003).

\bibitem{wien2k2} P. Blaha, K. Schwarz, G.K.H. Madsen, D. Kvasnicka, and J. Luitz,
 {\sc wien2k}, An Augmented Plane Wave+Local Orbitals Program for Calculating Crystal Properties
 (Karlheinz Schwarz, Techn. Universitaet Wien, Austria, 2001).


\bibitem{str1} T.Vogt and D. J. Buttrey,
  J. Solid State Chem. {\bf 123}, 186 (1996).

\bibitem{marzari} A.A. Mostofi, J.R. Yates, Y.-S. Lee, I. Souza, D. Vanderbilt, 
 and N. Marzari,
 Comput. Phys. Commun.{\bf 178}, 685 (2008).

\bibitem{jan10} J. Kune\v{s}, R. Arita, P. Wissgott, A. Toschi,
 H. Ikeda, and K. Held,
 Comput. Phys. Commun. {\bf 181}, 1888 (2010).


\bibitem{werner} P. Werner and A.J. Millis, 
 Phys. Rev. B {\bf 74}, 155107 (2006).

\bibitem{hafer} H. Hafermann, K.R. Patton, and P. Werner, 
 Phys. Rev. B {\bf 85}, 205106 (2012).

\bibitem{augustinsky} P. Augustinsk\'y and J. Kune\v{s}, 
 Comput. Phys. Commun.{\bf 184}, 2119 (2013).


\bibitem{maxent} M. Jarrell and J. E. Gubernatis, 
 Phys. Rep. {\bf 269}, 133 (1996).


\bibitem{ldau1} V. I. Anisimov, F. Aryasetiawan, and A. I. Lichtenstein,
 J. Phys.: Cond. Matt. {\bf 9}, 767 (1997).

\bibitem{ldau2}  E. R. Ylvisaker, K. Koepernik, and W. E. Pickett,
 Phys. Rev. B {\bf 79}, 035103 (2009).

\bibitem{fll} M. T. Czyzyk and G. A. Sawatzky,
 Phys. Rev. B {\bf 49}, 14211 (1994).


\bibitem{screen1} V. I. Anisimov, Dm. M. Korotin, M. A. Korotin, A. V. Kozhevnikov, J. Kune\v{s}, 
A. O. Shorikov, S. L. Skornyakov, and S. V. Streltsov, 
J. Phys.:Condens. Matter {\bf 21}, 075602 (2009); M. Aichhorn, L. Pourovskii, V. Vildosola, 
M. Ferrero, O. Parcollet, T. Miyake, A. Georges, and S. Biermann,
Phys. Rev. B {\bf 80}, 085101 (2009). 


\bibitem{krasko1987} G. L. Krasko, 
 Phys. Rev. B {\bf 36}, 8565 (1987).


\bibitem{byczuk} K. Byczuk, M. Kollar, K. Held, Y.-F. Yang, I. A. Nekrasov, Th. Pruschke, and D. Vollhardt,
Nat. Phys. {\bf 3}, 168 (2007).


\bibitem{pavarini} E. Pavarini, S. Biermann, A. Poteryaev, A. I.  Lichtenstein, A.  Georges, 
 and O. K. Andersen,
 Phys. Rev. Lett. {\bf 92}, 176403 (2004).





\end{thebibliography}
\end{document}